\documentclass[final,5p,twocolumn]{elsarticle/elsarticle}

\usepackage{amsmath,amsfonts,amssymb,mathrsfs,color}
\usepackage{natbib}
\usepackage{soul}
\usepackage{ulem}
\usepackage[dvipsnames]{xcolor}
\usepackage{hyperref}

\definecolor{mmcol}{rgb}{0,0.65,0.05}
\newcommand{\be}{\begin{equation}}
\newcommand{\ee}{\end{equation}}
\newcommand{\bea}{\begin{align}}
\newcommand{\eea}{\end{align}}
\newcommand{\bfig}{\begin{figure}}
\newcommand{\efig}{\end{figure}}

\newcommand{\om}{\Omega_{\rm m}}
\newcommand{\ob}{\Omega_{\rm b}}
\newcommand{\odm}{\Omega_{\rm DM}}
\newcommand{\ode}{\Omega_\Lambda}
\newcommand{\ho}{H_0}
\newcommand{\cinf}{c_\infty}
\newcommand{\lcdm}{$\Lambda$CDM}
\newcommand{\udm}{$\Lambda$CDM+$\cinf$}
\newcommand{\ncdm}{$\Lambda$CDM+$\sum m_\nu$}
\newcommand{\ud}{\mathrm{d}} 
\newcommand{\planck}{\textit{Planck}}

\journal{Physics of the Dark Universe}

\begin{document}

\begin{frontmatter}

\title{Does Quartessence Ease Cosmic Tensions?}

\author[unito,infn,inaf,jbca]{Stefano Camera}
\address[unito]{Dipartimento di Fisica, Universit\`a degli Studi di Torino, Via P. Giuria 1, 10125 Torino, Italy}
\address[infn]{INFN -- Istituto Nazionale di Fisica Nucleare, Sezione di Torino, Via P. Giuria 1, 10125 Torino, Italy}
\address[inaf]{INAF -- Istituto Nazionale di Astrofisica, Osservatorio Astrofisico di Torino, Strada Osservatorio 20, 10025 Pino Torinese, Italy}
\address[jbca]{Jodrell Bank Centre for Astrophysics, The University of Manchester, Oxford Road, Manchester M13 9PL, UK}
\ead{stefano.camera@unito.it}

\author[lorentz]{Matteo Martinelli}
\address[lorentz]{Instituut-Lorentz, Leiden University, PO Box 9506, Leiden 2300 RA, The Netherlands}
\ead{martinelli@lorentz.leidenuniv.nl}

\author[aifa]{Daniele Bertacca}
\address[aifa]{Argelander-Institut f\"ur Astronomie, Auf dem H\"ugel 71, 53121 Bonn, Germany}
\ead{dbertacca@astro.uni-bonn.de}

\begin{abstract}
Tensions between cosmic microwave background observations and the growth of the large-scale structure at late times pose a serious challenge to the concordance \lcdm\ cosmological model. State-of-the-art data from \planck\ predicts a higher rate of structure growth than what preferred by low-redshift observables. Such tension has hitherto eluded conclusive explanations in terms of straightforward modifications to \lcdm, e.g.\ the inclusion of massive neutrinos or dynamical dark energy. Here, we investigate 'quartessence'---a single dark component mimicking both dark matter and dark energy---whose non-vanishing sound speed inhibits structure growth at late times on scales smaller than its corresponding Jeans' length. In principle, this could reconcile high- and low-redshift observations. We test this hypothesis with temperature and polarisation spectra from the latest \planck\ release, SDSS DR12 measurements of baryon acoustic oscillations and redshift-space distortions, and cosmic shear from KiDS. This the first time that a specific model of quartessence is applied to actual data. We show that, if we naïvely apply \lcdm\ nonlinear prescription to quartessence, the combined data sets allow for tight constraints on the model parameters. Apparently, quartessence alleviates the tension between the total matter fraction and late-time structure clustering, although the tension is actually transferred from to the quartessence sound speed parameter. However, this strongly depends upon information from nonlinear scales. Indeed, if we relax this assumption, quartessence models appear still viable. For this reason, we argue that the nonlinear behaviour of quartessence deserves further investigation and may lead to a deeper understanding of the physics of the dark Universe.
\end{abstract}

\end{frontmatter}

\section{Introduction}\label{sec:i}
The current concordance cosmological model owes its name, \lcdm, to the two most abundant constituents of the present-day Universe: the cosmological constant, $\Lambda$, and a (cold) dark matter component. The former, responsible for the late-time accelerated expansion of the cosmos, amounts to $\sim70\%$ of the total energy budget; the latter, whose gravitational pull shaped the cosmic large-scale structure (LSS), constitutes more than $85\%$ of all the matter in the Universe, and roughly a quarter of its total content \citep{Ade:2015xua}.

Despite the success of the \lcdm\ model, a comparison among recent data sets suggests that the agreement between theory and observations is not adamantine. CMB data by \planck\ \citep{Ade:2015xua} is in tension with low-redshift observations, e.g.\ galaxy clustering as inferred through redshift-space distortions (RSD) \citep{Macaulay:2013swa}, weak gravitational lensing \citep{Heymans:2012gg,Hildebrandt:2016iqg}, and galaxy cluster counts \citep{Ade:2015gva}. Summarising, if we extrapolate CMB data to late times following \lcdm\ prescriptions, we will expect a higher rate of structure growth than what favoured by low redshift probes of the LSS \citep{Raveri:2015maa, Joudaki:2016mvz}. Notably, more recent LSS data, as cosmic shear from the Dark Energy Survey \citep{Troxel:2017xyo}, seems to exhibit a lesser degree of discrepancy with CMB data, but this does not in itself solve the issue of the pre-existing tensions.

Various approaches to tackling this serious problem have been adopted, ranging from re-analyses of observational data in the attempt of assessing the effect of possible systematics \citep[e.g.][]{Spergel:2013rxa,Addison:2015wyg}, to extensions or modifications of the \lcdm\ model. Among the latter approach, there is the inclusion of a free parameter accounting for the total mass of neutrinos \citep[e.g.][]{Battye:2013xqa,Joudaki:2016kym}. It is known from particle physics that neutrinos have small but non-negligible masses, as well as it is known, in cosmology, that the presence of massive neutrinos during the formation of the LSS causes a damping of matter fluctuations on scales smaller than the neutrino free-streaming length---which, in turn, is related to their masses. Thence the idea of reconciling CMB and LSS data by including massive neutrinos, whose effect will be negligible in the early Universe, although suppressing structure growth at low redshifts. Unfortunately, as shown by \citet{Joudaki:2016kym}, this scenario turns out not to be viable, mostly because of degeneracies between neutrino masses and other cosmological parameters.

Another possible route is that of dark energy and modified gravity theories \citep[for comprehensive reviews see][]{2010deto.book.....A,Clifton:2011jh}. As the underlying nature of dark matter and $\Lambda$ remains utterly unknown, a plethora of models, either phenomenological or emanated from first principles, have been proposed as alternatives to \lcdm. In literature, many of such extended theories have been considered over the last decade. In particular, let us mention the following, recent analyses:
\begin{itemize}
\item Using results from the latest \planck\ release \citep{Ade:2015xua}, the Kilo-Degree Survey (KiDS) \citep{Hildebrandt:2016iqg} and local $\ho$ estimates \citep{Riess:2016jrr}, \citet{Joudaki:2016kym} considered the curvature, constant-$w$ dark energy, modified gravity models, and running of the spectral index. 
The authors found that the inclusion of a constant $w$ parameter and/or curvature $\Omega_K\neq 0$   is enough to solve the tension, as these extended models are weakly favoured over \lcdm.
\item \citet{Pourtsidou:2016ico} and \citet{An:2017crg} studied interacting dark matter and dark energy models, where they consider a non-gravitational coupling between cold dark matter and dark energy. They concluded that this model can reconcile the tension between CMB observations and structure growth inferred from cluster counts.
\end{itemize}

An interesting family of such unhortodox cosmologies treats $\Lambda$ and dark matter as two faces of the same entity, a `dark  component' that both drives the current accelerated cosmic expansion and is responsible for the growth of the LSS. A large variety of these models---often called `unified dark matter' or `quartessence', in analogy to quintessence dark energy---are based on adiabatic fluids or on scalar field Lagrangians. On this topic, pioneering studies were made by e.g.\ \citet{Sahni:1999qe, Kamenshchik:2001cp, Bilic:2001cg} and have been further developed in Refs~\citep{Scherrer:2004au, Giannakis:2005kr, Bertacca:2007cv,Bertacca:2008uf}. For a recent review see \citet{Bertacca:2010ct} and subsequent works \citep{Bertacca:2010mt, Camera:2010wm, Bertacca:2011in,Camera:2012sf, Raccanelli:2012gt,Leanizbarrutia:2017afj}. The distinctive feature of this class of models is the existence of pressure perturbations in the rest frame of the quartessence, effectively originating a Jeans' length below which the growth of density inhomogeneities is impeded and the evolution of the gravitational potential is characterised by an oscillatory and decaying behaviour.

\section{Quartessence models}\label{sec:udm}
In this work we consider a particular class of quartessence models where a classical scalar field $\varphi$ with non-canonical kinetic term accounts for both dark matter a cosmological constant term in the background. We investigate the family of scalar-field quartessence models described by the following Lagrangian \citep{Bertacca:2008uf,Bertacca:2010ct}
\begin{equation}
\mathscr L_\mathrm{Q}=f(\varphi)g(X)-V(\varphi)\label{L_phi},
\end{equation}
where $g(X)$ is a Born-Infeld type kinetic term \citep{Born:1934gh} and
\begin{align}
f(\varphi)&=\frac{\Lambda \cinf}{1-{\cinf}^2}
\cosh(\xi\varphi) \nonumber\\
&\times \left[\sinh(\xi\varphi)\left[1+\left(1-{\cinf}^2\right)\sinh^2(\xi\varphi)\right]\right]^{-1} \\
V(\varphi)&=\frac{\Lambda}{1- {\cinf}^2}
\left[1+\left(1-{\cinf}^2\right)\sinh^2\left(\xi\varphi\right)\right]^{-1}\nonumber\\
&\times \left[\left(1-{\cinf}^2\right)^2\sinh^2\left(\xi\varphi\right)+2(1-{\cinf}^2)-1\right]
\end{align}
with $\xi=\sqrt{3/[4(1-{\cinf}^2)]}$, and $\cinf$ a free parameter. This Lagrangian can be thought as a field theory generalisation of the Lagrangian of a relativistic particle \citep{Padmanabhan:2002sh,Abramo:2003cp,Abramo:2004ji}.

The most important thing to bear in mind here is that these models are indistinguishable from \lcdm\ at background level. Furthermore, there is only one additional parameter with respect to \lcdm, as in the case of massive neutrinos. This parameter, $\cinf$, is related to the effective sound speed of quartessence, $c_{\rm s}^2$, and represents its asymptotic value at $t\to\infty$. 
Specifically, the speed of sound evolves with redshift according to \citep{Bertacca:2008uf}
\begin{equation}
c_{\rm s}^2(z)=\frac{\ode\cinf^{2}}{\ode+(1-\cinf^{2})\odm(1+z)^3}\label{c_s-udm}.
\end{equation}
Here, we `interpret' $\ode$  and $\odm$ as the present-day densities of an effective cosmological constant and a cold dark component, respectively. Note that this model recovers \lcdm\ for $\cinf=0$. In other words, the \lcdm\ parameter space is an hypersurface of the higher-dimensional parameter space of this family of quartessence models; we shall thus hereafter use the notation \lcdm\ and \udm. Similarly, we shall later refer to a cosmological model with free neutrino masses as \ncdm.

In quartessence, the evolution of the gravitational potential of the LSS is determined by the background and perturbation evolution of quartessence alone. This implies that the appearance of a sound speed significantly different from zero at late times will correspond to an effective Jeans length below which quartessence does clustering is inhibited \citep{Bertacca:2007cv}. To give a flavour of this effect, the quartessence sound speed of Eq.~\eqref{c_s-udm} with, say, $\cinf=10^{-3}$, determines a typical Jean's length for perturbations $\lambda_{\rm J}=16,1.6$ and $0.29\,h^{-1}\mathrm{Mpc}$ at $z=0,1$ and $2$, respectively. This quantity is directly proportional to $\cinf$ \citep[see][for further details on quartessence Jean's length]{Piattella:2009kt}. In turn, this will cause a strong evolution in time of the gravitational potential \citep{Camera:2009uz,Camera:2010wm}.

Then, for scales $k$ smaller than the cosmological horizon and redshift $z < z_{\rm rec}\simeq 1000$, we have
\begin{equation}
\delta(k,z)= T_{\rm Q}(k,z)\delta_{\rm m}(k,z),\label{eq:delta_UDM}
\end{equation}
where $\delta$ is the quartessence density contrast, $\delta_{\rm m}$ is the matter density perturbation in \lcdm\ and $T_{\rm Q}$ is the transfer function for the quartessence component. We adopt the approximate functional form
\begin{equation}
T_{\rm Q}(g) = 2^{5/8}\Gamma\left(\frac{13}{8}\right)g^{-5/8}J_{5/8}(g),\label{eq:T_UDM}
\end{equation}
where
$g[k,\eta(z)] = k\int_{\eta_{\rm rec}}^{\eta}\ud \eta'\,c_{\rm s}(\eta')$
and $\eta$ is the conformal time. \citet{Piattella:2011fv} prooved that the error produced by this approximation is almost negligible.\footnote{In particular, the absolute differences between the exact numerical result for $T_{\rm Q}$ and the fitting function in Eq.~\eqref{eq:T_UDM} is $\lesssim$1\%.}

\begin{figure}
\centering
\includegraphics[width=\columnwidth]{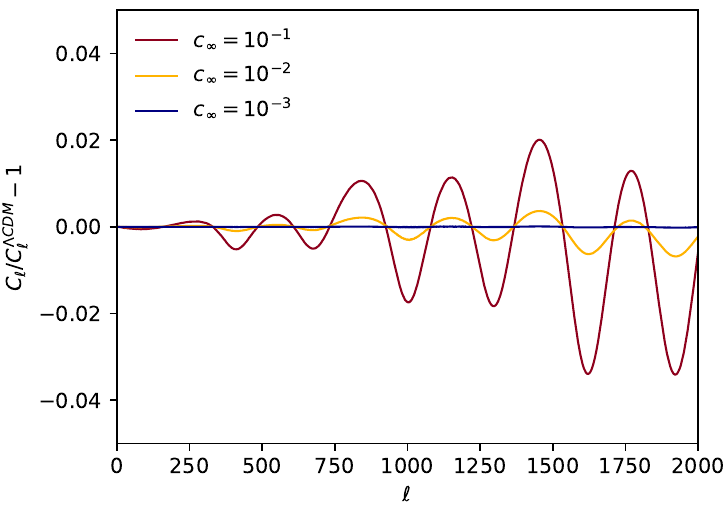}
\caption{Relative difference between the CMB temperature spectrum in \lcdm\ and in the case of a non-vanishing $\cinf$.}\label{fig:cmb}
\end{figure}

It is important to notice that, albeit negligible at the time of recombination, the presence of quartessence will also impact observations of the CMB. Indeed, CMB photons are lensed by cosmological structures---an effect leading to a smoothing of the acoustic peaks of CMB spectra. Since the larger $\cinf$ the more suppressed the growth of matter perturbations, increasing $\cinf$ values therefore lead to lower amplitudes of the CMB lensing potential \citep{Camera:2009uz,Camera:2010wm}. In turn, \udm\ CMB spectra differ from the standard \lcdm\ prediction at angular scales corresponding to the acoustic peaks, as shown in Fig.~\ref{fig:cmb}. This effect is described here for the first time, but let us remark that the impact of quartessence peculiar clustering on CMB lensing power spectrum had already been already investigated in \citet[][see their Fig.~4]{Camera:2009uz}.

\section{Methodology}\label{sec:iv}
\subsection{Observables and data sets}
As a result of the modified evolution of density fluctuations when $\cinf\neq0$, several cosmological observables differ from the \lcdm\ expectation. For a start, we can look at galaxies' peculiar velocities and gravitational lensing distortions induced by the intervening LSS on light emitted by distant sources. To find signatures of a non-vanishing $\cinf$, we therefore analyse currently available measurements of baryon acoustic oscillations (BAO) and RSD from the Sloan Digital Sky Survey (SDSS) DR12 \citep{Beutler:2016ixs,Satpathy:2016tct}, and the correlation functions of weak lensing cosmic shear from KiDS \citep{Hildebrandt:2016iqg,Joudaki:2016kym}. We name the combination of BAO, RSD and cosmic shear data set as `LSS'.\footnote{As a matter of fact, BAO is a `geometrical' observable, as it only probes the background cosmic expansion. Nonetheless, we included it among the LSS data set, alongside the other low-$z$ probes.} (Note that in the DR12 SDSS data set the tension with \planck\ data reported by \citet{Macaulay:2013swa} has disappeared.)

Thanks to the effect of quartessence on CMB lensing, described at the end of the previous section, we can also use CMB data to scrutinise the viability of \udm\ models. Hence, we employ temperature and polarisation spectra from the latest \planck\ release \cite{Aghanim:2015xee}. In particular, we employ the \texttt{plik} TT+TE+EE likelihood together with \texttt{low-$\ell$} polarisation data. Furthermore, to be consistent with the latest results from the \planck\ collaboration \cite{Adam:2016hgk}, we include a Gaussian prior on the optical depth to reionisation, $\tau=0.058\pm0.012$.

On top of this CMB data, we also include in our analysis the CMB lensing power spectrum obtained through quadratic estimators by \citet{Ade:2015zua}, when combining also with the LSS data set. However, as we shall see in Fig.~\ref{fig:1dcon} (left panel), the impact of this additional data set, although helping in constraining $\cinf$, does not affect significantly the results. For this reason, and given the slight discrepancies of the results inferred on the amplitude of lensing effect between the CMB spectra and the lensing quadratic estimator, we opt for not including it when estimating the tension between CMB and LSS, as well as when we calculate the model selection estimators.

\subsection{Data analysis}
Summarising, our aim is to state whether or not the non-standard clustering of quartessence models can ease the CMB-LSS tension. First, we proceed to quantify the constraints on $\cinf$ obtained from CMB and LSS data. For this purpose, we sample the standard \lcdm\ 6 parameters space: the baryon and cold dark matter physical densities, $\omega_{\rm b}=\ob h^2$ and $\omega_{\rm DM}=\odm h^2$; the sound horizon at the last scattering surface, $\theta_\ast$; the amplitude and tilt of the primordial spectrum of scalar perturbations, $A_s$ and $n_s$; and the optical depth to recombination, $\tau$. To these parameters, we add $\cinf$.

For comparison, we also analyse the case of massive neutrinos, \ncdm, for which we sample the same cosmological parameter space adding one dimension related to a free value for the sum of neutrino masses, $\sum m_\nu$ \citep[cfr][]{Battye:2013xqa,Battye:2014qga}. The parameter space is sampled assuming flat priors for all the parameters and using Monte-Carlo Markov Chains (MCMC) and a Gelman-Rubin convergence diagnostic implemented in the publicly available code \texttt{CosmoMC} \citep{Lewis:2002ah,Lewis:2013hha}. This is interfaced with a version of the public code \texttt{CAMB} \citep{Lewis:1999bs,Howlett:2012mh}, modified in order to account for $\cinf$ by including the modifications shown in Eqs~(\ref{eq:delta_UDM})-(\ref{eq:T_UDM}) in the dark matter transfer function of \texttt{CAMB}. To test the robustness of our results, in \ref{sec:app} we also explore the impact of our choice of sampling on $\cinf$. In particular, we make a change of variables, moving from $\cinf$ to $\log_{10}\cinf^2$ and assuming, again, flat priors on all parameters.

It is often common in the literature, for extensions of \lcdm\ model affecting the growth of cosmic structures, to remove from the analysis scales where nonlinear effects are relevant \citep[e.g.][]{Ade:2015rim,Joudaki:2016kym}, not to rely on nonlinear modelling based on \lcdm\ numerical simulations. Here, however, we deal with $\cinf$ values that do not lead to significant deviations from the standard \lcdm\ behaviour at the scales probed by the data \citep[see also][]{Camera:2010wm}.\footnote{As we shall see in \S~\ref{sec:v}, a value already excluded at 2$\sigma$ such as $\cinf=10^{-3}$ leads to a less than 2\% deviation in the matter power spectrum at $k=0.1\,h\mathrm{Mpc}^{-1}$.} Therefore, we decide to include such scales, using the corrections to the linear power spectrum computed by the \texttt{HMCODE}~\citep{2015MNRAS.454.1958M}.

\section{Results and discussion}\label{sec:v}
Table~\ref{tab:res} shows the results obtained with different data-set combinations. As in the recent literature \citep{Hildebrandt:2016iqg,Joudaki:2016kym}, we also quote constraints on the derived parameter $S_8=\sigma_8 \sqrt{\Omega_{\mathrm{m}}/0.3}$, which is the combination of $\sigma_8$, the rms mass fluctuations on a scale of $8\,h^{-1}\mathrm{Mpc}$, and the total matter fraction $\om=h^{-2}(\omega_{\rm b}+\omega_{\rm DM})$, mainly constrained by current weak lensing data. The latter is often used as a proxy to exemplify the CMB-LSS tension.
\begin{table*}
\caption{\label{tab:res}Marginalised values and 1-$\sigma$ errors for $S_8$ and 2-$\sigma$ upper bound on $\cinf$ and $\sum{m_\nu}$ for \planck, LSS and \planck+lens+LSS.}
\vspace{0.1cm}
\centering
\begin{tabular}{lccccc}
\hline
&& \planck & LSS &  \planck+LSS & \planck+lens+LSS\\
\hline
\hline
\lcdm & & & & & \\
\hline
&$S_8$                 & $0.850\pm 0.017$           & $0.741\pm 0.026$              & $0.817\pm 0.012$           & $0.815\pm 0.010$\\
\hline
\hline
\udm & & & & & \\
\hline
&$\cinf$               & $<5\times10^{-3}$          & $<0.6\times10^{-3}$           & $<0.5\times10^{-3}$        & $<0.5\times10^{-3}$\\
&$S_8$                 & $0.719^{+0.15}_{-0.046}$   & $0.736^{+0.029}_{-0.026}$     & $0.814\pm 0.012$            & $0.810^{+0.012}_{-0.010}$\vspace{0.1cm}\\
\hline
\hline
\ncdm & & & & & \\
\hline
&$\sum{m_\nu}\ [{\rm eV}]$   & $<0.59$                      & $<2.9$                        & $<0.21$                      & $<0.21$\\
&$S_8$                 & $0.837^{+0.021}_{-0.019}$    & $0.743\pm 0.027$              & $0.809\pm 0.014$           & $0.811\pm 0.011$\vspace{0.1cm}\\
\hline
\end{tabular}
\end{table*}

\begin{figure*}
\centering
\includegraphics[width=\columnwidth]{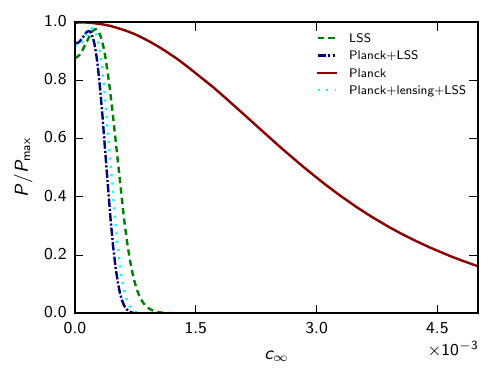}\includegraphics[width=\columnwidth]{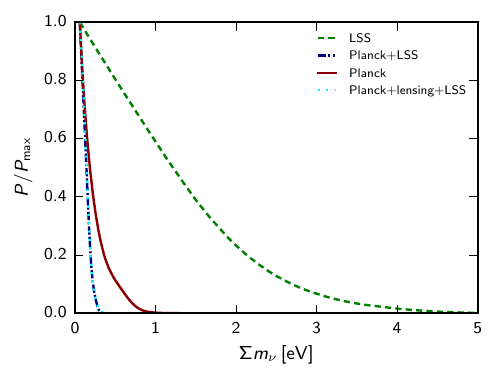}
\caption{Posterior distributions for $\cinf$ (left panel) and $\sum m_{\nu}$ (right panel), showing constraints from \planck\ (red, solid lines), LSS observables (green, dashed lines) and the combination of the two (blue, dot-dashed lines).}\label{fig:1dcon}
\end{figure*}

\begin{figure*}
\centering
\includegraphics[width=\columnwidth]{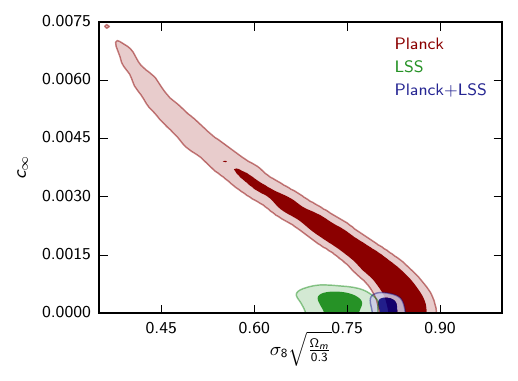}\includegraphics[width=\columnwidth]{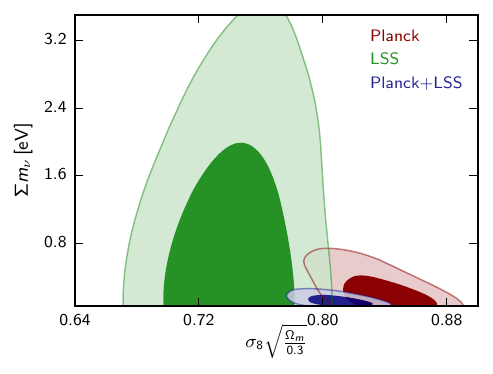}
\caption{2D contour plots showing the degeneracy between $S_8$ and $\cinf$ or $\sum m_\nu$ (left or right panel, respectively), for the various combinations of data sets examined, with darker (lighter) areas depicting 68.3\% (95.5\%) joint marginal bounds.}
\label{fig:2Dres}
\end{figure*}

These results highlight how $\cinf$ is strongly constrained by LSS observables, whilst \planck\ data alone, which is only affected by quartessence through CMB lensing, allows for larger values of the parameter, as it can be seen in Figs~\ref{fig:1dcon} and \ref{fig:2Dres}. This translates into much broader bounds on $S_8$ from CMB data and with a lower mean value for this parameter combination with respect to the standard \lcdm\ bound ($S_8=0.719^{+0.15}_{-0.046}$), a scenario opposite to that of \ncdm, where the neutrino mass is strongly constrained by CMB.

We emphasise that, albeit the allowed range of values for quartessence speed-of-sound parameter $\cinf$ may at a first glance seem fine tuned, this is in fact not the case. 
Indeed, let us point out that, at linear order, it is possible to recast quartessence Lagrangians as effective axion-like Lagrangians in the Thomas-Fermi approximation \citep{Chavanis:2011uv}, where the effective mass of the axion changes with the time (this mapping is currently under investigation and it is left to future work).

\begin{figure}
\centering
\includegraphics[width=\columnwidth]{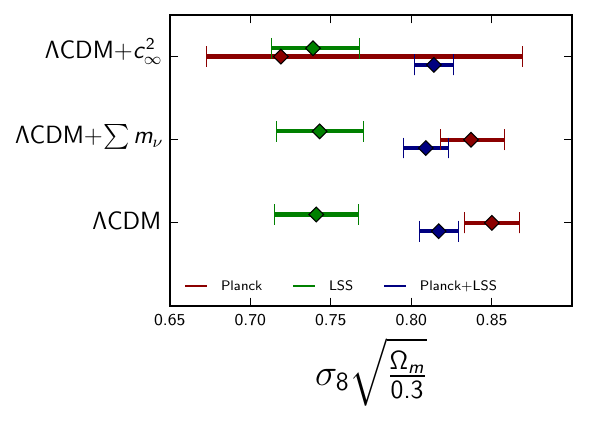}
\caption{Marginal 1$\sigma$ bounds on $S_8=\sigma_8(\om/0.3)^{0.5}$ obtained by LSS (green lines) and CMB (red lines) data sets, and their combination (blue line) in the three cosmological scenarios under investigation.}
\label{fig:tension}
\end{figure}

As a result of the peculiar quartessence dynamics, the CMB-LSS tension can be eased by quartessence with $\cinf\neq0$ due to its effect on $S_8$, as shown in Fig.~\ref{fig:2Dres}. In the left panel, it is easy to see that larger values of $\cinf$ imply a smaller $S_8$, thus reconciling high- and low-redshift observations. This can be also appreciated by quantifying the $S_8$ tension with the estimator proposed by Refs~\citep{Hildebrandt:2016iqg,Joudaki:2016kym}
\begin{equation}
 T(S_8)=\frac{\left|S_8^{\rm CMB}-S_8^{\rm LSS}\right|}{\sqrt{\sigma^2(S_8^{\rm CMB})+\sigma^2(S_8^{\rm LSS})}},
\end{equation}
where we remind the reader that by `CMB' and `LSS' we indicate that the parameter constraint is obtained from a \planck\ or KiDS+BAO+RSD analysis, respectively. For the \udm\ model considered here, we obtain $T(S_8)=0.3$, therefore quartessence strongly eases the $3.5\sigma$ tension we find when comparing CMB and LSS data in standard \lcdm. Another way to see this is shown in Fig.~\ref{fig:tension}, where the 1$\sigma$ error intervals on $S_8$, marginalised over all the other parameters, are shown for \lcdm, \udm\ and \ncdm, with red, green and blue lines for CMB, LSS and their combination, respectively. Our results on neutrinos slightly differ from those by \citet{Joudaki:2016kym}, because we use more recent data compared to them. Similarly, more recent data sets, as weak lensing cosmic shear from the first year of data taking of the Dark Energy Survey, appear in better agreement with \planck, but nonetheless give $T(S_8)=2$ \citep[see][Table~III]{Troxel:2017xyo}.

\begin{figure*}
\centering
\includegraphics[width=\columnwidth]{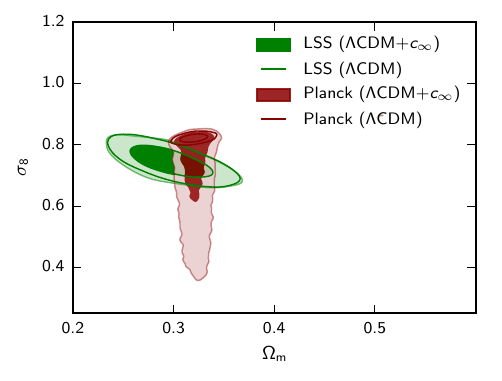}\includegraphics[width=\columnwidth]{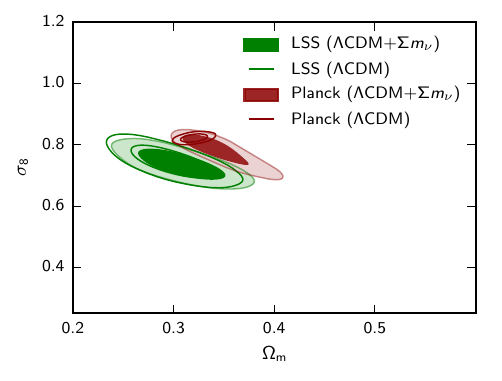}
\caption{Marginal 1- and 2-$\sigma$ joint error contours in the $\Omega_{\rm m}-\sigma_8$ plane obtained by CMB and LSS measurements (red and green contours, respectively). The plot shows the results in $\Lambda$CDM (filled contours) and in \udm\ and \ncdm\ (left and right panel, respectively).}
\label{fig:tritension}
\end{figure*}

Figure \ref{fig:tritension} shows the 2D joint marginal error contours for $\om$ and $\sigma_8$ in standard \lcdm\ and for both \udm\ and \ncdm. Interestingly, even though both $\sum m_\nu$ and $\cinf$ are able to ease the $\sigma_8$ tension (although to a much different extent), their impact is substantially different, thus we find that a joint analysis \lcdm +$\cinf+\sum{m_\nu}$ would not significantly change the constraints.

Finally, we explore another way to quantify these result, namely we perform a model selection analysis by computing the deviance information criterion (DIC) \cite{RSSB:RSSB12062}, in order to assess which of the considered models is favoured by the data. For a given model, the DIC is defined as 
\begin{equation}\label{eq:dic}
\mathrm{DIC}\equiv\chi^2_{\rm eff}(\hat{\vartheta})+2p_{\rm D},
\end{equation}
with $\chi^2_{\rm eff}(\hat{\vartheta})= -2\ln{\mathcal{L}(\hat{\theta})}$, $\hat{\vartheta}$ the parameters vector at the maximum likelihood and
\begin{equation}
p_{\rm D}=\overline{\chi^2_{\rm eff}(\vartheta)}-\chi^2_{\rm eff}(\hat{\vartheta}),
\end{equation}
where an average taken over the posterior distribution is implied by the bar. The term $p_{\rm D}$ accounts for the complexity of the model, balancing the improvement brought on the goodness of fit, $\chi^2_{\rm eff}$, by the introduction of additional parameters.

We compute $\Delta\mathrm{DIC}$ for the two extended models, using the DIC of \lcdm\ as a reference; given the definition of Eq.~\eqref{eq:dic}, a negative (positive) value of $\Delta\mathrm{DIC}$ will show that the extended model is favoured (disfavoured) by the data over \lcdm. Combining CMB and LSS data sets we find that in \udm\ $\Delta\mathrm{DIC}=-0.21$; although this results is better than what found for \ncdm\ ($\Delta\mathrm{DIC}\approx1$ when combining CMB and LSS), no preference for quartessence over the standard \lcdm\ cosmology is found. To state this, we assume thresholds of $-5$ and $-10$ for moderate and strong preference of the extended model over \lcdm, as discussed in \citet{Joudaki:2016kym}.

The poor evidence in favour of quartessence is due to the fact that, whilst CMB results leave quite some freedom for the $\cinf$ parameter, thus inabling to ease the tension, the LSS data set tightly constraints this parameter (with constraining power coming mostly from nonlinear scales). Hence, when combining the two data-sets to evaluate the DIC, the LSS dominates and does not allow $\cinf$ to assume the values able to ease the tension. As a result, the DIC does not significantly improve with respect to \lcdm. Therefore, we conclude that the effect of the \udm\ model on the growth of structure decreases significantly the tension on the $S_8$ parameter: $0.3\sigma$ for \udm\ vs $3.4\sigma$ for \lcdm. However, the comparison of the two models, quantified through the DIC, highlights that the available data does not favour quartessence over \lcdm\ in a statistically significant manner.

This result might seem puzzling at a first glance, given the significant reduction of the tension on $S_8$. However, assessing the tension focusing only on a single parameter might be misleading as, moving from $\Lambda$CDM to \udm, the tension on $S_8$ might have been moved in other parts of the parameter space. This can be quantified exploiting the DIC estimator to assess the concordance of the two different data sets \citep[see][]{Joudaki:2016kym},
\begin{equation}
2\log{I}=\mathcal{G}({\rm CMB},{\rm LSS}),
\end{equation}
with 
\begin{equation}
\mathcal{G}({\rm CMB},{\rm LSS})={\rm DIC}_{{\rm CMB}\cup {\rm LSS}}-{\rm DIC}_{\rm CMB}-{\rm DIC}_{\rm LSS}.
\end{equation}
This quantity estimates the concordance of the data sets, evaluating, through the DIC estimator, the ratio between the evidence of the two data sets being described by the same set of parameters, and the evidence of the two being described by different parameter sets. We find $\Delta\log{I}\approx 0.5$, meaning a not substantial evidence in favour of \udm. This implies a negligible difference between $\Lambda$CDM and \udm. Hence, we can conclude that widening the parameter space from \lcdm\ to \udm\ does not significantly improve the concordance of CMB and LSS data sets, despite an apparent easing of the tension.

Let us emphasise that these conclusions depend on the estimators used to quantify the tension, namely $T(S_8)$, and to compare the models, viz.\ $\Delta\textrm{DIC}$. Our results justify further investigations of quartessence models, which we leave for future work, in order to assess this dependence using other tension estimators \citep[e.g.][]{Charnock:2017vcd, Raveri:2018wln} and more refined model comparison techniques, such as the computation of the Bayesian evidence for the extended model \citep{Heavens:2017hkr}. In any case, we believe that the reconciliation of early- and late-Universe observations attained by quartessence models makes them worth deeper investigations, in particular for what concerns their nonlinear behaviour, either with dedicated numerical $N$-body simulations or through nonlinear perturbation theory approaches.

\section*{Acknowledgements}
We wish to thank an anonymous reviewer, whose insightful questions helped improving the presentation of our results. We thank Anna Bonaldi, Antonaldo Diaferio, Shahab Joudaki and Tom Kitching for valuable support. SC is supported by the Italian Ministry of Education, University and Research (MIUR) through Rita Levi Montalcini project `\textsc{prometheus} -- Probing and Relating Observables with Multi-wavelength Experiments To Help Enlightening the Universe's Structure', and by the `Departments of Excellence 2018-2022' Grant awarded by MIUR (L.\ 232/2016). SC also acknowledges support from ERC Starting Grant No.\ 280127. MM is supported by the Foundation for Fundamental Research on Matter (FOM) and the Netherlands Organization for Scientific Research / Ministry of Science and Education (NWO/OCW). DB is supported by the Deutsche Forschungsgemeinschaft through the Transregio 33, The Dark Universe.

\appendix\section{Effects of parameter sampling}\label{sec:app}
Throughout the paper, we have assumed in our MCMCs a linear sampling for $\cinf$ with a flat prior over the whole allowed range. However, as shown also in Fig.~\ref{fig:cmb}, this parameter is in principle spanning several orders of magnitude; therefore, one could also think to sample $\log_{10}{\cinf^2}$ rather than $\cinf$, allowing the MCMC to probe lower values of this parameter better. Such values we expect to be favoured by the data.

\begin{figure}
\centering
\includegraphics[width=\columnwidth]{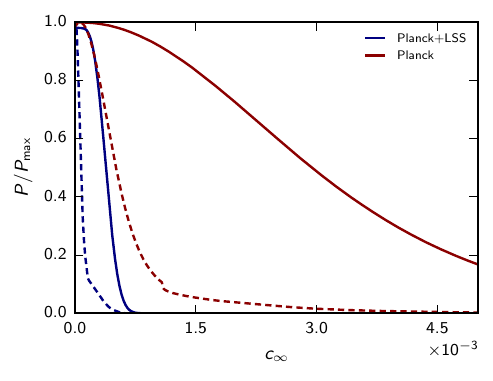}
\caption{Posterior distributions for $\cinf$ showing constraints from \planck\ (red lines) and the combination \planck +LSS (blue lines). Solid lines represent constraints obtained with a flat prior on $\cinf$ whilst dashed lines refer to the case where the flat prior is adopted on $\log_{10}{\cinf^2}$.}
\label{fig:sampling}
\end{figure}

Choosing a flat prior on $\log_{10}{\cinf^2}$ would however bias the results towards low values of $\cinf$. Indeed, if one computes the marginalised posterior probability on a parameter $\theta$ of a $d$-dimensional set $\Theta$, $P(\theta)$, as a function of the likelihood $\mathcal{L}(\Theta)$ and priors $\Pi(\Theta)$, this will read
\begin{equation}\label{eq:margpost}
 P(\theta)\propto\int\ud^{d-1}\Theta\,\mathcal{L}(\Theta)\Pi(\Theta).
\end{equation}
Therefore, changing parameterisation $\theta\rightarrow\zeta(\theta)$ implies that a prior that was flat on $\theta$ will not be so for $\zeta$ in general. In turn, the priors in Eq.~\eqref{eq:margpost} will change according to
\begin{equation}
 \Pi(\zeta)=\Pi(\theta)\frac{\ud\theta}{\ud\zeta}.
\end{equation}
As shown in \cite{Giannantonio:2009gi}, a logarithmic sampling produces a negative tilt of the prior on the physical parameter, in our case effectively down-weighting large $\cinf$ values. This is indeed what we find if we re-analyse our cases sampling $\log_{10}{\cinf^2}$ rather than $\cinf$, as shown in Fig.~\ref{fig:sampling}.

Given this result, we chose in this paper to sample the linear $\cinf$ parameter in order not to bias our results towards \lcdm. However, as it is particularly evident in the \planck\ alone case, the difference between the two choices can reach pretty significant levels and solving this issue will require more sensitive data, able to constrain better this parameter and to limit the range allowed for it.

\bibliographystyle{elsarticle/elsarticle-num-names}
\bibliography{biblio}
\biboptions{sort&compress}

\end{document}